# Acoustic guiding and subwavelength imaging with sharp bending by sonic crystal

**Bo Li, Ke Deng[*], and Heping Zhao**

Department of Physics, Jishou University, Jishou 416000, Hunan, China

A sharp bending scheme for the self-collimation of acoustic waves is proposed by simply truncating the sonic crystals. An all-angle and wide-band 90°-bending wave guide is demonstrated with nearly perfect transmissions for Gaussian beams at a wide range of incident angles. A 90°-bended imaging for a point source with a subwavelength resolution of $0.37\lambda_0$ is also realized by the proposed structure. These results will find applicability in the manipulation of acoustic waves by sonic crystals.

PACS numbers: 63.20.-e, 43.35.+d, 43.58.+z

Sonic crystals (SCs) provide an exhilarating platform for the manipulation of acoustic waves. Suitable designing of a SC gives the phenomenon called self-collimation, by which a beam of acoustic wave can propagate in SCs with almost no diffraction along a definite direction [1,2]. This interesting phenomenon, also reported in Photonic crystals (PCs) [3-5], affords a promising way to control the flow of waves, thus it has been utilized for various wave-functional applications, such as the wave beaming [6,7], subwavelength imaging [8,9] and wave guiding [10].

In the case of wave guiding, self-collimation allows varying over a wide range of incident angles and still maintaining a narrow range of propagations along with a low-loss transmission. This type of waveguide does not depend on physical channels to confine waves, thus it gains many advantages over conventional line-defect based wave guides, for which introducing of physical boundaries in SCs or PCs leads to extra coupling loss and Fresnel reflection loss [3,5]. Moreover, profited from a long flat equifrequency contour (EFC), evanescent modes of a source placed at the input port of such waveguides can also be canalized to the output port [8,9], thus self-collimation waveguides possess a promising prospect in subwavelength applications. However, such wave guide is suffered from the intrinsic inability of self-collimation to efficiently bend and redirect waves. Actually, guiding waves around sharp corners is also not feasible for classical total-internal-reflection (TIR) based wave guides. From this point of view, realization of high-efficiency sharp bending with low-loss transmission is of significant importance in wave controlling applications.

[*]Corresponding author, e-mail address: dengke@jsu.edu.cn.

In this paper, we propose a high-efficiency sharp bending technique for the self-collimation of acoustic waves by simply truncating the SC. As demonstrated by the rigid multiple-scattering theory (MST) [11,12], low-loss transmissions of a 90° bending wave guide for Gaussian beams with a wide range of incident angles can be achieved over wide frequency ranges. Furthermore, this technique is applied to realize a 90° bended subwavelength imaging for a point source with a resolution of $0.37\lambda_0$. These results are expected to shed some light on various applications of wave manipulation.

We consider a two-dimensional SC composed of rubber-coated tungsten rods arranged as a square lattice in water. The outer and inner radii of rubber are $0.429a$ and $0.398a$ respectively, where $a$ is the lattice constant. The material parameters are mass density $\rho = 19.3\times10^3\,\text{kg m}^{-3}$, longitudinal wave velocity $C^{\text{L}} = 5.09\times10^3\,\text{m s}^{-1}$ and transverse wave velocity $C^{\text{T}} = 2.8\times10^3\,\text{m s}^{-1}$ for tungsten; $\rho = 1.3\times10^3\,\text{kg m}^{-3}$, $C^{\text{L}} = 0.2\times10^3\,\text{m s}^{-1}$ and $C^{\text{T}} = 0.04\times10^3\,\text{m s}^{-1}$ for rubber; $\rho = 1.0\times10^3\,\text{kg m}^{-3}$ and $C^{\text{L}} = 1.49\times10^3\,\text{m s}^{-1}$ for water.

Band structure of the SC is shown in Fig. 1(a). Frequencies considered here are around 0.115. Figure 1(b) presents several EFCs around 0.115. These EFCs are practically flat and are much longer compared to those of water in the first Brillouin zone (as an example, the dashed circle around $\Gamma$ point denotes the EFC of water at 0.115). This implies that if we cut the SC along $\Gamma\text{M}$ direction and put a source in water near the interface, all propagating spatial harmonics of the source and a wide range of evanescent ones will refract into SC's eigenmodes with wave vectors closed to $q$ [see Fig. 1(b)] and then undergo a self-collimated propagation along the $\Gamma\text{M}$ direction around the frequency 0.115 [8,9]. In addition, as these eigenmodes have wave vector components parallel to the $\Gamma\text{X}$ direction [$q_{//}$ in Fig. 1(b), for example] that lie outside the EFCs of water, a SC-water interface along $\Gamma\text{X}$ direction would behave as a TIR mirror for these self-collimated beams [13,14].

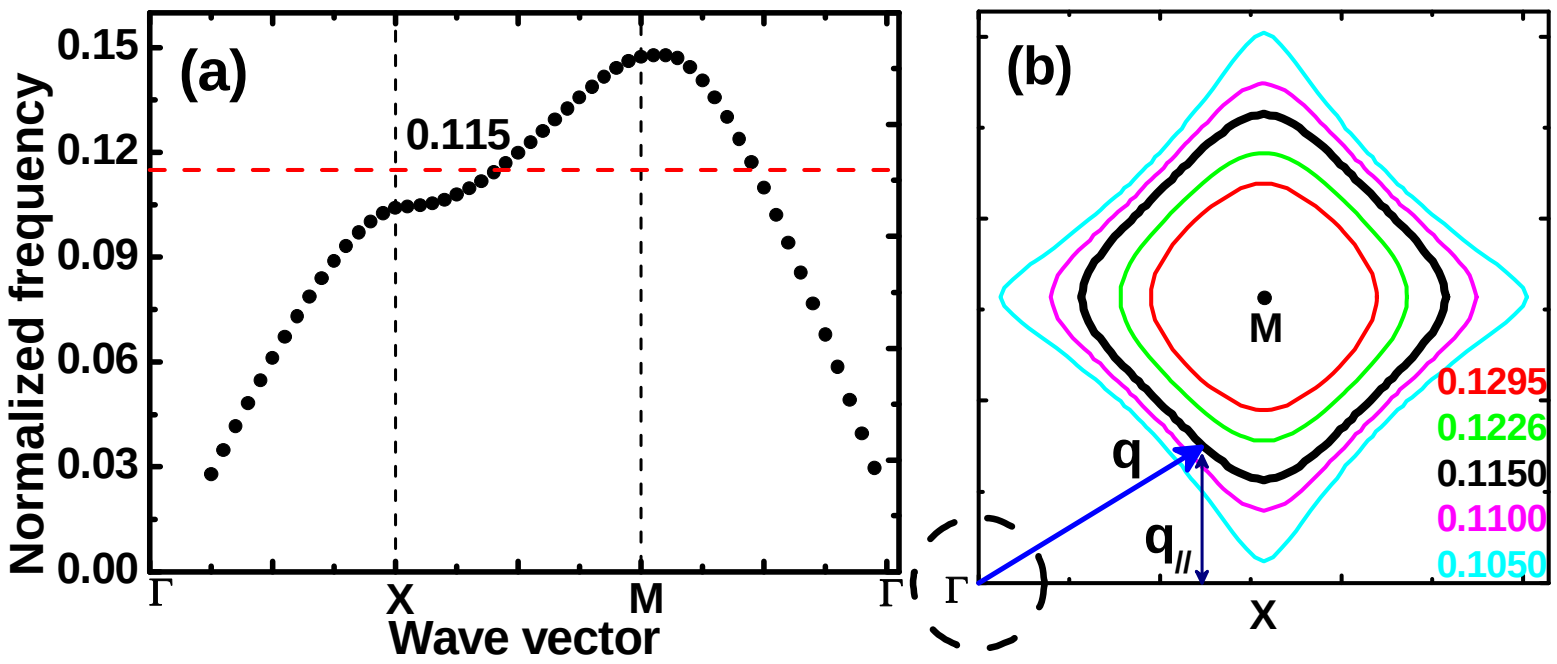

FIG. 1 (Color online) (a) The lowest band for SC considered here. The frequency is normalized to $2\pi c / a$, where $c$ is the acoustic speed in water. (b) Several EFCs around 0.115. The dashed circle is the EFC of water at frequency 0.115.

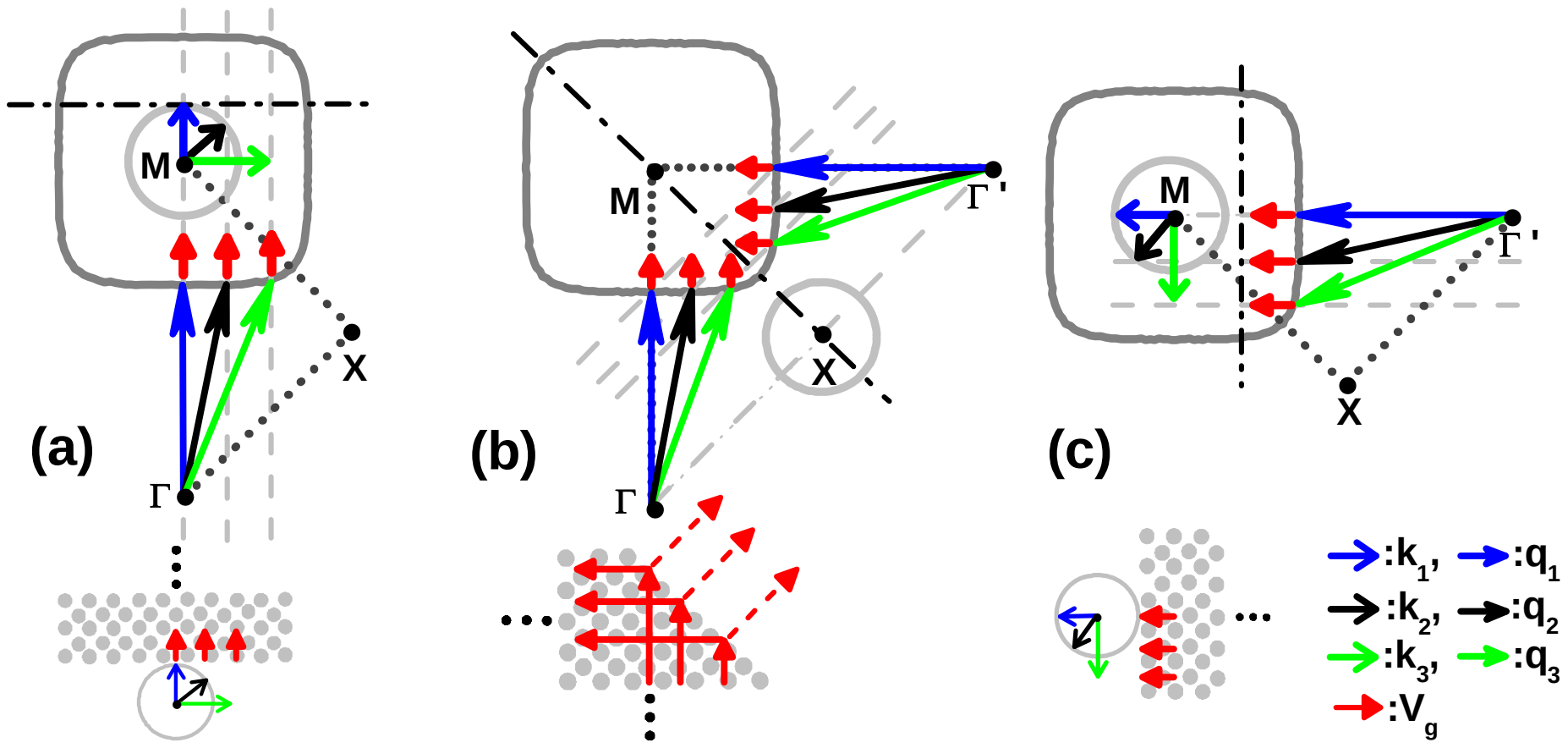

FIG. 2 (Color online) (a) Refraction of three wave vectors from a source into SC through a ΓM directed interface. (b) Total bending of collimated waves in SC at a ΓX directed interface. (c) Refraction of the bended waves into a source-like image at another ΓM directed interface.

To make these two conclusions more clearly, we give a detailed analysis for the case of 0.115 in Fig. 2. Here the gray square-like curve around M point and the light gray cycle represents EFC of SC and water, respectively. The dash dot lines denote the orientation of SC-water interfaces. Firstly, we consider two propagating waves $k_1$ and $k_2$ incident upon a ΓM directed SC-water interface from, for example, a source in water with $0°$ and $40°$, as illustrated by Fig. 2(a). Their refracted modes can be determined by the conservation-law of wave vector component parallel to the (ΓM) interface. By drawing the k-conservation lines (dash lines in Fig.2) in the direction perpendicular to the interface, we find their refracted modes ($q_1$ and $q_2$ starting from $\Gamma$ point). These refracted rays propagate in SC along the same directions as the group velocity $V_g = \nabla_q [\omega(q)]$ which are perpendicular to SC's EFC, as indicated by the red thick arrows in Fig. 2(a). In this case one sees that an all-angle self-collimation can be achieved. In Fig. 2(b) we consider another SC-water interface oriented along $\Gamma X$ direction. When $q_1$ and $q_2$ encountered such an interface, reflection and refraction will happen in SC and into water respectively. The reflected and refracted modes can also be obtained by the conservation-law of parallel wave vector [14]. As indicated in Fig. 2(b), by drawing the k-conservation lines, we can find the reflected modes (represented as $q_1$ and $q_2$ starting from $\Gamma'$ point) which are self-collimated along the horizontal direction as shown in Fig. 2. This implies that the collimated modes are bent by 90° in SC after reflection. We note that the k-conservation lines in Fig. 2(b) lie outside of the EFC of water, i.e., $q_1$ and $q_2$ have wave vector components parallel to the $\Gamma X$ direction that are lager than the wave vector in water. This means that a TIR will occur and this bending efficiency is 100%, since according to the conservation of parallel wave vector, refracted waves in water will have evanescent components normal to the ($\Gamma X$) interface, as denoted by the dashed arrows in the lower panel of Fig. 2(b). Combining the above analyses, we conclude that an all-angle wave-guiding with 90° bending can be realized by the proposed SC structure with proper truncations.

On the other hand, we consider an evanescent mode of the source as denoted by $k_3$ in Fig. 2. We point out that, due to the particular shape of EFCs, the above analysis to $k_1$ and $k_2$ holds exactly for $k_3$. As a matter of fact, in this case all spatial harmonics of the source between $[-k_3, k_3]$ can refract into SC with almost the same group velocity $V_g$ and undergo a self-collimation, and then in the similar way undergo a TIR at the $\Gamma X$ interface as discussed above. After bending, these collimated waves then refract into a source-like image when a $\Gamma M$ directed interface is introduced again, as shown in Fig. 2(c). Because a substantial range of the evanescent modes of the point source has been canalized across the SC, the formed image will have subwavelength resolution. Thus it can be seen that the guiding and bending channel discussed above also possesses the abilities to transport some subwavelength information of the source and to image. According to these above analysis, an all-angle guiding and 90˚-bending scheme with subwavelength imaging ability can be expected.

To verify the above discussions, we introduce into SC a truncation along ΓX direction as shown in Fig. 3(a), and vertically illuminate the structure with a Gaussian beam at frequency 0.115. Pressure field distribution is shown in Fig. 3(a) which exhibits a remarkable self-collimation and bending effect. Figure 3(b) gives the bending efficiency for vertically illuminated Gaussian beams with different frequencies. One sees that near 90% efficiency can be accomplished over the entire frequency range in which self-collimation occurs.

In Fig. 4 we show the case of oblique incidence. Pressure field distribution for a Gaussian beam at frequency 0.115 with incident angle $\theta = 40^o$ is shown in Fig. 4(a), which also exhibits a remarkable self-collimation and bending effect. Figure 4(b) gives the bending efficiency for Gaussian beams at frequency 0.115 with different incident angles. The results clearly show that high-efficiency can be achieved when incident angle is smaller than 40˚. For large incident angles, the efficiency falls sharply as the result of rapidly increased reflections at the input port of SC structure.

Figure 5 gives results for the imaging of a point source. The source is placed pretty close to ($\sqrt{2}a/2$ away from) the under surface of a truncated SC lens. Figure 5(a) gives the pressure field distribution and Figure 5(b) gives the transverse pressure field distribution at the imaging plane. The full width at half maximum of the image is approximately $0.37\lambda_0$, exhibiting a subwavelength resolution.

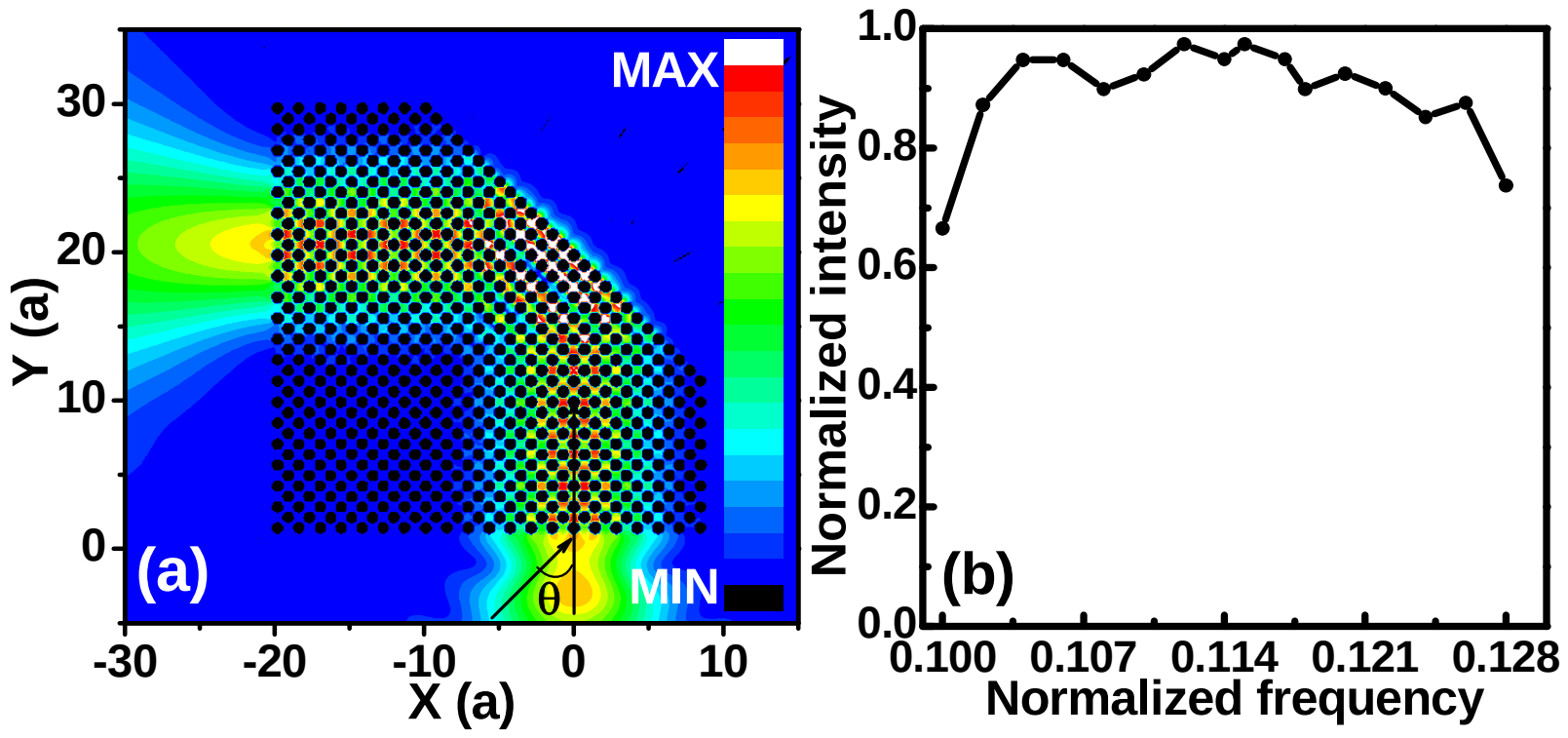

FIG. 3. (Color online) (a) Pressure field distribution for a vertically illuminated Gaussian beam at frequency 0.115. (b) Bending efficiency for vertically illuminated Gaussian beams with different frequencies.

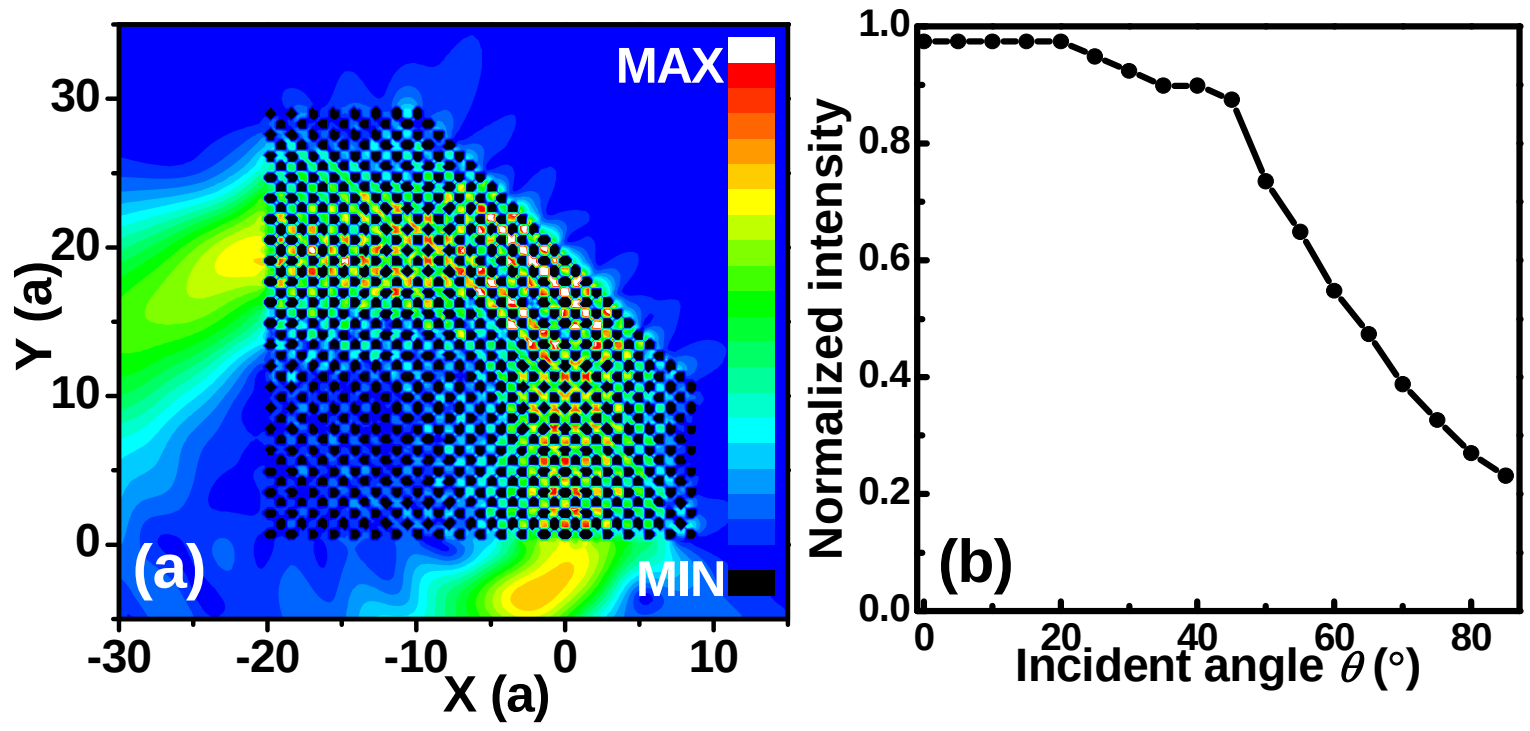

FIG. 4. (Color online) (a) Pressure field distribution for a Gaussian beam at frequency 0.115 with incident angle $\theta = 40^{o}$. (b) Bending efficiency for Gaussian beams at frequency 0.115 with different incident angles.

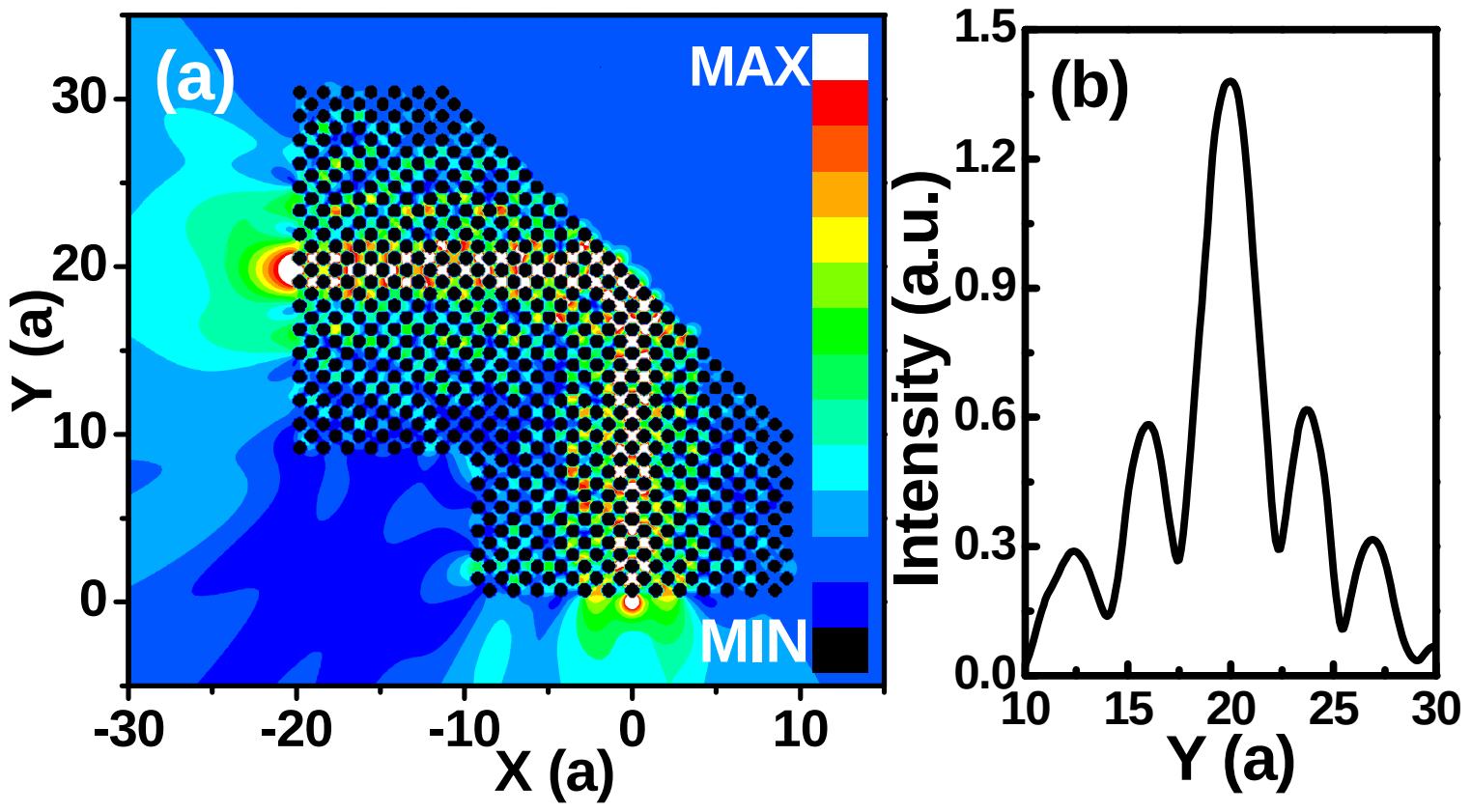

FIG. 5. (Color online) (a) Pressure field distribution of a point source placed $\sqrt{2}a/2$ away from the under surface of a truncated SC lens. (b) Transverse field distribution at the imaging plane of (a).

To summarize, we have proposed an easy-to-implement scheme for the sharp bending of self-collimated acoustic waves in SCs. An all-angle 90°-bending wave guide with a wide bandwidth was demonstrated, by which nearly perfect transmissions were achieved for Gaussian beams with a wide range of incident angles. In addition, this scheme was utilized to design a lens structure for a point source to realize a 90°-bended subwavelength imaging. With such a structure, propagating and evanescent spatial harmonics of the source can refract into the eigenmodes of SC and then be canalized along collimated or bended directions as one desires. These results will find significant applicability, particularly in steering acoustic waves by SCs and in various subwavelength applications.

This work is a Project supported by Hunan Provincial Natural Science Foundation of China (Grant No. 09JJ6011 and 11JJ6007), and supported by Natural Science Foundation of Education Department of Hunan Province, China (Grant No. 08A055).